\documentclass[cameraready]{Interspeech}

\title{Dissecting Sensitivity to Training Language in Self-Supervised Speech Learning Using Neural Audio Codec Tokens}

\author[affiliation={1}, orcid=0009-0006-3530-0830, equalcontribution]{Daigo}{Takizawa}
\author[affiliation={1}, orcid=0000-0003-4385-7170, equalcontribution]{Tomohiko}{Nakamura}
\author[affiliation={2}, orcid=0000-0002-5358-1844, equalcontribution]{Samuele}{Cornell}
\author[affiliation={2}, orcid=0000-0002-3251-3084, equalcontribution]{William}{Chen}
\author[affiliation={1}, orcid=0000-0001-6506-2796, equalcontribution]{Satoru}{Fukayama}
\author[affiliation={2}, orcid=0000-0002-5970-8631, equalcontribution, correspondingauthor]{Shinji}{Watanabe}

\address{
    $^1$ National Institute of Advanced Industrial Science and Technology  (AIST) , Japan \\
    $^2$ Carnegie Mellon University, USA
}

\email{daigo.takizawa@aist.go.jp, s.fukayama@aist.go.jp, shinjiw@ieee.org}


\usepackage{comment}
\usepackage{graphicx}
\usepackage{cleveref}
\usepackage{booktabs}
\usepackage{multirow}
\usepackage{xurl}
\usepackage{subcaption}
\usepackage{cite}

\crefname{figure}{Fig.}{Figs.}
\newcommand{\ER}{\text{ER}}
\newcommand{\ERwav}{\widetilde{\ER}_{d}}

\begin{document}

\maketitle

\begin{abstract}
Neural audio codecs (NACs) have become popular for obtaining speech representations as discrete tokens. Beyond compression, discrete tokens can be used to train self-supervised learning (SSL) models. Such models, referred to as codec-based SSL models, reduce data storage and computational cost, enabling scalable SSL pre-training. However, their language sensitivity remains unclear. When the language changes, codec-based SSL models may require retraining, which undermines their efficiency. In this paper, we present a systematic analysis of language sensitivity by varying either the NAC training language or the SSL pre-training language while keeping the other fixed. Experimental results show that downstream performance is insensitive to the NAC training language but strongly dependent on the SSL pre-training language. These findings suggest that a single NAC can be reused across languages, while aligning the SSL pre-training language with the target language is crucial.
\end{abstract}

\section{Introduction} \label{sec:intro}
Self-supervised learning (SSL) is widely used to build universal feature extractors for speech tasks such as automatic speech recognition (ASR) and speech emotion recognition (SER)~\cite{Mohamed2022IEEEJSTSP,baevski2020wav2vec, hsu2021hubert}.
Despite this versatility, SSL typically requires unlabeled yet massive amounts of speech data and long training schedules, limiting the scalability of pre-training in storage and computation.
To realize scalable SSL, cost-efficient pre-training is needed to retain broad downstream utility~\cite{Chen2023Interspeech,Yang2023ASRU,Lugo2024EACL,Liu2024SLT,shi2024espnet,tseng2025codec2vec}.

A promising approach for cost-efficient pre-training is to use compact representations produced by a neural audio codec (NAC).
A NAC compresses waveforms into code sequences and can reconstruct the corresponding waveforms from them~\cite{Mousavi2025TMLR}.
Beyond compression, these codec units (discrete tokens) have been shown to retain rich acoustic and linguistic information and to be directly usable for speech processing tasks without waveform reconstruction~\cite{shi2024espnet}. 
In the SSL literature, models have been trained directly on such discrete tokens, achieving reductions in storage and computational cost during pre-training~\cite{tseng2025codec2vec}.
We refer to SSL pre-training with discrete-token inputs as \emph{codec-based SSL}.

In codec-based SSL, language sensitivity can arise from both the NAC discretizer and the SSL pre-training stage, potentially confounding their contributions.
On the SSL side, prior work has shown that downstream performance degrades when the pre-training language is misaligned with the target language~\cite{ashihara2023exploration,chou2023asru_hokkien,wang2024how_to_learn_new_language}.
On the NAC side, reconstructed-waveform quality varies across languages~\cite{altwlkany2025language}, while NACs can generalize to languages unseen during pre-training in speech reconstruction and text-to-speech tasks~\cite{wang2026neural}.
However, the extent to which the \emph{NAC training language} affects codec-based SSL performance remains underexplored.
This distinction has practical implications: requiring NAC retraining for each target language would diminish its cost advantages.

\begin{figure}[t]
\centering
\includegraphics[width=0.85\columnwidth,clip]{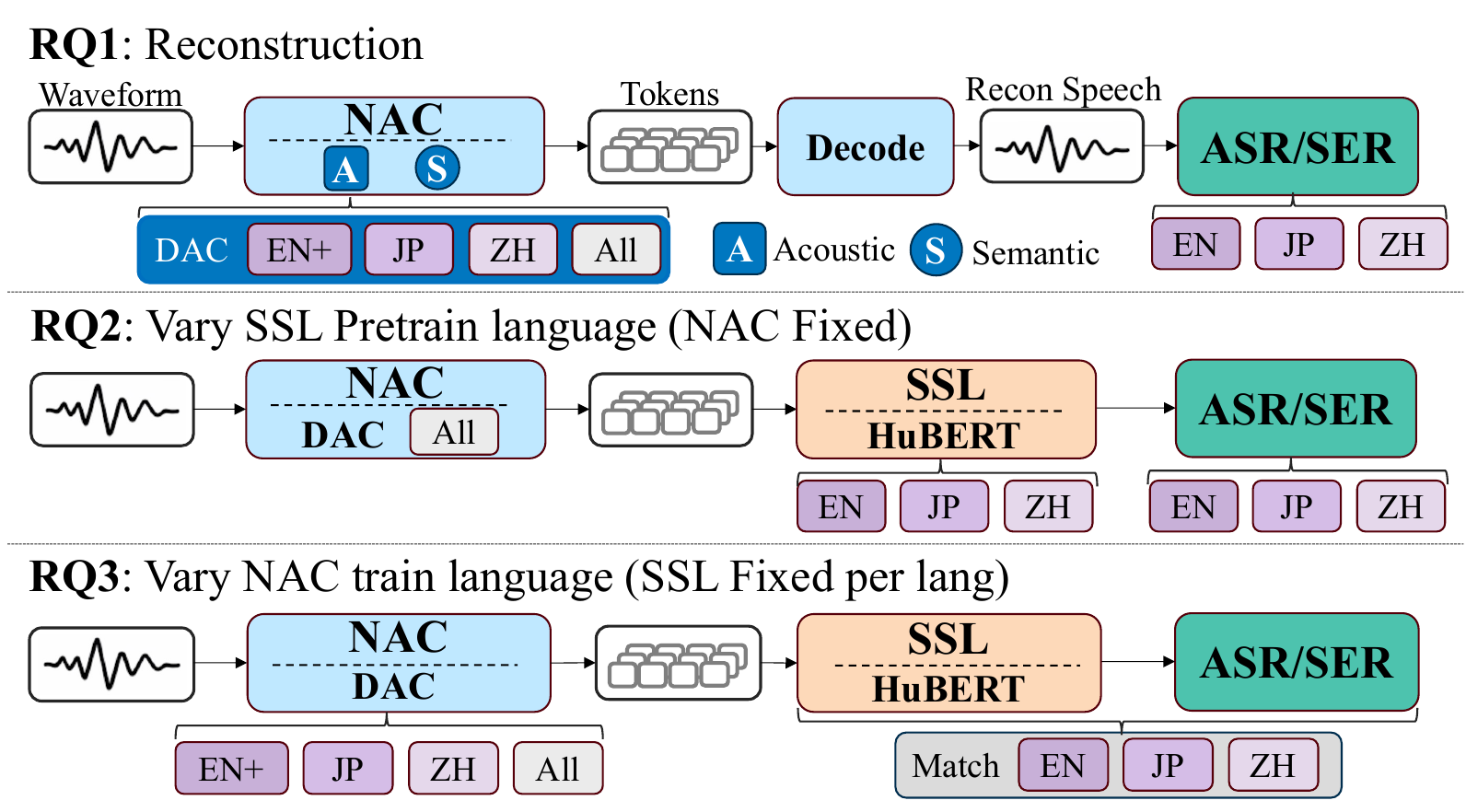}
\caption{Overview of our controlled experimental design to disentangle language sensitivity in codec-based SSL by decoupling the NAC and SSL pre-training stages. RQ1 evaluates downstream performance on NAC-reconstructed waveforms, RQ2 varies the SSL pre-training language with a fixed NAC, and RQ3 varies the NAC training language with matched SSL pre-training.}
\label{fig:overview}
\end{figure}

In this paper, we conduct a controlled study assessing whether NAC retraining is necessary under language shifts, following the staged design in Fig.~\ref{fig:overview}. 
We first examine NAC language sensitivity on reconstructed speech to characterize its cross-lingual behavior (RQ1). 
We then analyze language sensitivity in codec-based SSL using discrete tokens as model inputs (RQ2 and RQ3). 
Specifically, we address the following three research questions:
\begin{enumerate}[
    label=\textbf{RQ\arabic*:},
    leftmargin=*,
    labelsep=0.5em,
    align=left,
    topsep=0pt,
    partopsep=0pt,
]
\item {How does the NAC training language affect downstream performance on reconstructed waveform?}
\item {In codec-based SSL, how does the SSL pre-training language affect downstream task performance?}
\item {When the SSL pre-training language is fixed to the downstream language in codec-based SSL, how does the NAC training language affect downstream performance?}
\end{enumerate}

\section{Related Work}
Codec-based SSL has recently gained attention as a cost-efficient alternative to waveform-based SSL.
ESPnet-Codec provided a unified benchmark for training and evaluating NACs and offered early empirical studies of codec-based SSL~\cite{shi2024espnet}.
Codec2Vec showed that codec-based SSL can substantially reduce storage and computational costs while maintaining performance comparable to waveform-based SSL~\cite{tseng2025codec2vec}.
However, these studies are English-centric and do not assess cross-lingual differences or language sensitivity.
Our paper is the first to provide a systematic analysis of language sensitivity for codec-based SSL.

Language sensitivity has been observed for both waveform-based SSL and NACs.
For waveform-based SSL, low-resource ASR studies report performance degradation under language mismatch between pre-training and downstream task~\cite{ashihara2023exploration,chou2023asru_hokkien,wang2024how_to_learn_new_language}.
For NACs, reconstruction quality can vary across languages~\cite{altwlkany2025language}, while NACs can generalize to unseen languages in speech reconstruction and text-to-speech (TTS) tasks~\cite{wang2026neural}.
By contrast, we quantify how NAC reconstruction impacts downstream \emph{recognition} across languages and isolate the role of NAC training language via controlled retraining.

\section{Evaluation Setup} \label{sec:protocol}

\subsection{Experimental Design} \label{sec:exp_design}
The experiments are designed to decouple language sensitivity in codec-based SSL across the NAC and the SSL pre-training stage (\cref{fig:overview}).
For RQ1, we evaluate ASR/SER on NAC-reconstructed waveform using multiple publicly available NACs and Descript Audio Codec (DAC)~\cite{kumar2023high} retrained under different training-language conditions.
For RQ2, we fix the NAC and vary only the SSL pre-training language in codec-based SSL to quantify its effect on downstream tasks.
For RQ3, we fix the SSL pre-training language to match the downstream language and vary only the NAC training language to isolate its effect within codec-based SSL.

Across all settings, we evaluate two downstream tasks, ASR and SER, in English, Japanese, and Chinese.
These languages are chosen due to the availability of large-scale public datasets for both pre-training and evaluation, as well as their diverse linguistic characteristics. 
We use ASR and SER as complementary downstream tasks, capturing linguistic/phonetic information and acoustic/prosodic information, respectively.

\begin{table}[t]
\centering
\caption{Overview of training data used at each stage.
Dataset abbreviations: non-speech (music and general audio followed the public data sources in~\cite{kumar2023high}),
LL-10h (Libri-Light~\cite{kahn2020libri} 10h), LTVS-100h (LaboroTVSpeech~\cite{ando2021construction} 100h),
WS-100h (WenetSpeech subset-S~\cite{zhang2022wenetspeech}), AS1 (AISHELL-1~\cite{bu2017aishell}).}
\label{tab:data_overview}
\footnotesize
\setlength{\tabcolsep}{3pt}
\begin{tabular}{l l l r @{}r}
\toprule
\multicolumn{5}{l}{\textbf{(A) NAC Training}}\\
\midrule
Cond. & \multicolumn{2}{l}{Dataset (Lang.)} & \multicolumn{2}{r}{Hours} \\
\midrule
EN+  & \multicolumn{2}{l}{Speech (EN/FR/DE/RU/ES) + non-speech} & 1056 & \\
JP   & \multicolumn{2}{l}{In-house Broadcast TV data (JP)} & 1056 & \\
ZH   & \multicolumn{2}{l}{WenetSpeech subset-L (ZH)~\cite{zhang2022wenetspeech} + non-speech} & 1056 & \\
All  & \multicolumn{2}{l}{Speech (EN+/JP/ZH) + non-speech} & 1056 & \\
\midrule
\multicolumn{5}{l}{\textbf{(B) SSL Pre-training}}\\
\midrule
Lang. & \multicolumn{2}{l}{Dataset} & \multicolumn{2}{r}{Hours} \\
\midrule
EN & \multicolumn{2}{l}{LibriSpeech~\cite{panayotov2015librispeech}} & 960 & \\
JP & \multicolumn{2}{l}{In-house Broadcast TV data} & 4821 & \\
ZH & \multicolumn{2}{l}{WenetSpeech subset-L~\cite{zhang2022wenetspeech}} & 7173 & \\
\midrule
\multicolumn{5}{l}{\textbf{(C) Downstream}}\\
\midrule
Task & Lang. & Dataset(s) & \multicolumn{2}{r}{Hours} \\
\midrule
    & EN & LL-10h & 10 & \\
ASR & JP & LTVS-100h / CSJ~\cite{maekawa03_sspr} / COJADS~\cite{cojads} & 100 / 516 / 60 & \\
    & ZH & WS-100h / AS1 & 100 / 150 & \\
\midrule
    & EN & IEMOCAP~\cite{busso2008iemocap} & $\sim$12 & \\
SER & JP & JTES~\cite{takeishi2016construction} & $\sim$24 & \\
    & ZH & EmotionTalk (EmoTalk)~\cite{sun2025emotiontalk} & $\sim$24 & \\
\bottomrule
\end{tabular}
\end{table}

\subsection{Downstream Tasks}
\label{sec:downstream}

\noindent\textbf{ASR.}
ASR experiments were conducted using ESPnet~\cite{watanabe2018espnet}.
For each language, models were trained and evaluated separately on the corresponding datasets summarized in Table~\ref{tab:data_overview}.
All datasets followed the official splits used in ESPnet, except for the splits for LaboroTVSpeech~\cite{ando2021construction} where only the first \SI{100}{\hour} of the official training split was used, and the splits for COJADS~\cite{cojads} where evaluation data were selected to ensure regional balance with the remaining data used for training and validation.
A Conformer model was trained for each dataset with SSL features as input, without an external language model.
Layer-wise SSL representations were combined by a trainable weighted sum.
Word error rate (WER) was reported for English and character error rate (CER) for Japanese and Chinese.

\noindent\textbf{SER.}
SER experiments followed the SUPERB emotion recognition setting~\cite{yang2021superb} using the datasets listed in Table~\ref{tab:data_overview}.
IEMOCAP~\cite{busso2008iemocap} and JTES~\cite{takeishi2016construction} were evaluated with a \num{5}-fold cross-validation, while with EmotionTalk (EmoTalk)~\cite{sun2025emotiontalk} we used its official split.
Four emotion categories (happy, angry, sad, neutral) were used for all datasets. Average recall (AR) was reported: weighted AR for JTES and unweighted AR for IEMOCAP and EmoTalk due to the class imbalance.

\begin{table*}[t]
\centering
\caption{
ASR and SER performance on NAC-reconstructed waveforms across NAC types and languages (RQ1).
For LL-10h and WS-100h, X/Y denotes (test\_clean/test\_other) and (test\_net/test\_meeting), respectively.
For CSJ, results are averaged over eval1–eval3.
CoV is computed across languages as defined in Section~\ref{sec:metric}.
Dataset abbreviations are defined in Table~\ref{tab:data_overview}.
“Multi” denotes multilingual training data reported in the original papers.
EN+, JP, ZH, and All denote DAC training-language conditions defined in Section~\ref{sec:comp_nac_train_lang}.
}
\label{tab:rq1_results}

{%
\footnotesize
\setlength{\tabcolsep}{2.9pt}
\begin{tabular}{l c|c ccc cc|ccc|c}
\toprule
\multicolumn{2}{c|}{} &
\multicolumn{6}{c|}{\textbf{ASR} [\%] $\downarrow$} &
\multicolumn{3}{c|}{\textbf{SER} [\%] $\uparrow$} &
\\
\cmidrule(lr){3-8}\cmidrule(lr){9-11}

\multicolumn{2}{c|}{\textbf{Train. Cond.}} &
\multicolumn{1}{c}{EN} &
\multicolumn{3}{c}{JP} &
\multicolumn{2}{c|}{ZH} &
\multicolumn{1}{c}{EN} &
\multicolumn{1}{c}{JP} &
\multicolumn{1}{c|}{ZH} &
\multicolumn{1}{c}{\textbf{CoV} [\%] $\downarrow$} \\
\cmidrule(lr){1-2}
\cmidrule(lr){3-3}\cmidrule(lr){4-6}\cmidrule(lr){7-8}
\cmidrule(lr){9-9}\cmidrule(lr){10-10}\cmidrule(lr){11-11}\cmidrule(lr){12-12}

Representation & Lang.
& LL-10h
& LTVS-100h & CSJ & COJADS
& WS-100h & AS1
& IEMOCAP & JTES & EmoTalk
& ASR / SER \\
\midrule

\multicolumn{12}{l}{\textbf{Top line}} \\
\midrule
Waveform & --
& 10.2 / 17.7 & 13.5 & 4.3 & 34.7
& 13.7 / 19.2 & 4.4
& 65.11 $\pm$ 3.1 & 78.47 $\pm$ 4.4 & 77.88
& - \\
\midrule

\multicolumn{12}{l}{\textbf{Publicly available NACs (\Cref{sec:comp_existing_nacs})}} \\
\midrule
DAC~\cite{kumar2023high} & Multi
& \textbf{10.5 / 19.1} &\textbf{14.3} & \textbf{4.5} & \textbf{37.9}
& \textbf{15.3 / 22.9} & \textbf{4.8}
& 65.43 $\pm$ 1.8 & 77.52 $\pm$ 4.2 & \textbf{77.77}
& \textbf{3.38 / 2.22} \\
EnCodec~\cite{defossezhigh} & Multi
& 10.9 / 20.1 & 14.9 & 4.9 & 39.9
& 16.2 / 25.1 & 5.2
& \textbf{66.27 $\pm$ 1.9} & 73.61 $\pm$ 3.7 & 75.74
& 4.61 / 9.62 \\
SpeechTokenizer~\cite{xin2024speechtokenizer} & EN
& 11.9 / 23.6 & 20.3 & 6.4 & 52.3
& 26.0 / 51.3 & 8.5
& 64.07 $\pm$ 2.2 & 75.21 $\pm$ 2.6 & 74.73 
& 23.72 / 4.99 \\
X-Codec~\cite{ye2025codec} & EN
& 10.8 / 19.8 & 16.8 & 5.1 & 43.9
& 20.6 / 30.6 & 5.5
& 65.27 $\pm$ 2.0 & 77.12 $\pm$ 3.2 & 75.07
& 12.04 / 5.05 \\
PAST~\cite{har2025past} & EN
& 10.9 / 20.7 & 19.0 & 5.8 & 46.5
& 22.6 / 43.8 & 9.1
& 65.62 $\pm$ 1.6 & \textbf{77.62 $\pm$ 3.5} & 73.66 
& 25.06 / 8.10 \\
\midrule

\multicolumn{12}{l}{\textbf{Effect of NAC Training Language (\Cref{sec:comp_nac_train_lang})}} \\
\midrule
\multirow{4}{*}{DAC (ours)} & EN+
& 10.4 / \textbf{18.6} & \textbf{14.0} & \textbf{4.5} & 36.9
& \textbf{14.6} / 21.6 & 4.7
& 65.58 $\pm$ 2.3 & 77.32 $\pm$ 4.6 & 78.14
& 2.15 / 3.08 \\
& JP
& 10.4 / \textbf{18.6} & 14.2 & \textbf{4.5} & 36.9
& 14.7 / 21.7 & 4.8
& 65.40 $\pm$ 2.3 & \textbf{77.87 $\pm$ 4.3} & 78.36
& 2.53 / \textbf{2.09} \\
& ZH
& 10.4 / 18.9 & \textbf{14.0} & \textbf{4.5} & 37.0
& 14.7 / 21.9 & 4.7
& 65.18 $\pm$ 2.6 & 76.97 $\pm$ 4.0 & 77.98
& 2.12 / 3.37 \\
& All
& \textbf{10.3 / 18.6} & 14.1 & \textbf{4.5} & \textbf{36.8}
& 14.7 / \textbf{21.4} & \textbf{4.6}
& \textbf{65.71 $\pm$ 2.3} & 77.67 $\pm$ 3.7 & \textbf{78.45}
& \textbf{1.92} / 2.78 \\
\bottomrule
\end{tabular}%
}

\end{table*}

\subsection{Metric for Language Sensitivity} \label{sec:metric}
Language sensitivity was quantified using the coefficient of variation (CoV), a scale-free measure for comparing variability among groups with different means. 
We compare the error rates divided by the ones with the baseline model, which takes waveform as input, so that the resulting ratio-scale values could be compared among different language settings.
Let $\mathcal{L}$ denote the set of languages and $\mathcal{D}_l$ the set of evaluation datasets in language $l \in \mathcal{L}$.
For each dataset $d \in \mathcal{D}_l$, the error rate $\ER_d$ was normalized by the corresponding waveform-input baseline error rate $\ERwav$ and averaged within each language; CoV was then computed across languages as
\begin{equation}
    r_l=\frac{1}{|\mathcal{D}_l|}\sum_{d \in \mathcal{D}_l}\frac{\ER_d}{\ERwav},
    \enspace
    \text{CoV}=\frac{\sqrt{\sum_{l \in \mathcal{L}}\left( r_l - \bar{r} \right)^2/|\mathcal{L}|}}{\bar{r}},
\end{equation}
where $\bar{r}=\sum_{l \in \mathcal{L}} r_l/|\mathcal{L}|$.
As for $\ER_d$, we used WER (English) or CER (Japanese/Chinese) for ASR; one minus AR for SER.

\section{Language Sensitivity in NAC-Reconstructed Waveforms}

\subsection{Comparison across NACs} \label{sec:comp_existing_nacs}

To answer RQ1, we examined language sensitivity on NAC-reconstructed waveforms using the setup in \Cref{sec:protocol}.
We evaluated five publicly available neural audio codecs with distinct training specifications: DAC~\cite{kumar2023high}, EnCodec~\cite{defossezhigh}, SpeechTokenizer~\cite{xin2024speechtokenizer}, X-Codec~\cite{ye2025codec}, and PAST~\cite{har2025past}\footnote{
We used publicly released checkpoints: \url{https://github.com/descriptinc/descript-audio-codec/releases/download/0.0.5/weights_16khz.pth} for DAC;
\url{https://huggingface.co/facebook/encodec_24khz} for EnCodec;
\url{https://huggingface.co/hf-audio/xcodec-wavlm-mls} for X-Codec;
\url{https://huggingface.co/OpenMOSS-Team/SpeechTokenizer/tree/main/speechtokenizer_hubert_avg} for SpeechTokenizer;
\url{https://huggingface.co/slprl/PAST/blob/main/PAST.th} for PAST.
}.
DAC and EnCodec were trained on multilingual data including Japanese and Chinese, while the other models were trained primarily on English data and incorporated auxiliary objectives such as SSL guidance or ASR-related supervision (see the respective papers for details).

The second block of \Cref{tab:rq1_results} summarizes ASR and SER performance across the publicly available NACs.
The first block reports results using the original waveforms.
The performances for cross-lingual conditions differed within NACs. 
For instance, DAC achieved the smallest CoVs for both ASR and SER among the evaluated models.
We therefore select DAC for subsequent analysis so that we do not need to consider the effect of using different NACs.

\subsection{Effect of Choosing Language for Training NAC} \label{sec:comp_nac_train_lang}
To examine the effect of using different languages for NAC training, we trained DAC under four conditions: \emph{EN+}, \emph{JP}, \emph{ZH}, and \emph{All} (see Table~\ref{tab:data_overview}).
In the EN+ condition, following~\cite{kumar2023high}, we used the same data sources but restricted Common Voice to English to avoid Japanese and Chinese speech.
For fair comparison, $1056$ hours of data were randomly sampled for each condition.
We used the official DAC implementation, and the training recipe was based on the official \SI{16}{\kilo\hertz} configuration\footnote{\url{https://github.com/descriptinc/descript-audio-codec/blob/main/conf/final/16khz.yml}.}, with \num{18} codebooks to improve reconstruction quality, increasing the bitrate from \SI{6}{kbps} to \SI{9}{kbps}.

The third block of \Cref{tab:rq1_results} reports the ASR and SER performance on reconstructed waveform for DACs trained under different language conditions.
The variation within cross-lingual conditions remained small, and we did not observe a systematic trend, even for the condition of \emph{All}.
These results demonstrated that the DAC training language has only a limited impact on downstream performance on reconstructed waveform, which is consistent with previous observations in TTS tasks~\cite{wang2026neural}.

\smallskip
\noindent \textbf{RQ1:}
Language sensitivity for training NAC is limited especially for acoustic tokenizers, and DAC is the most stable among the evaluated NACs.
Thus, the NAC training language is not a primary factor driving cross-lingual variation in downstream tasks when evaluated on NAC-reconstructed waveform.

\begin{table*}[t]
\centering
\caption{
ASR and SER performance when varying NAC and SSL training languages in codec-based SSL.
`Match' denotes the downstream language.
Other notation (X/Y, CoV) and dataset abbreviations follow Table~\ref{tab:data_overview}.
}
\label{tab:rq2_rq3_results}

{%
\footnotesize
\setlength{\tabcolsep}{2.9pt}
\begin{tabular}{cc|c ccc cc|ccc|c}
\toprule
\multicolumn{2}{c|}{} &
\multicolumn{6}{c|}{\textbf{ASR} [\%] $\downarrow$} &
\multicolumn{3}{c|}{\textbf{SER} [\%] $\uparrow$} &
\\
\cmidrule(lr){3-8}\cmidrule(lr){9-11}

\multicolumn{2}{c|}{\textbf{Train. Lang.}} &
\multicolumn{1}{c}{EN} &
\multicolumn{3}{c}{JP} &
\multicolumn{2}{c|}{ZH} &
\multicolumn{1}{c}{EN} &
\multicolumn{1}{c}{JP} &
\multicolumn{1}{c|}{ZH} &
\multicolumn{1}{c}{\textbf{CoV} [\%] $\downarrow$} \\
\cmidrule(lr){1-2}
\cmidrule(lr){3-3}\cmidrule(lr){4-6}\cmidrule(lr){7-8}
\cmidrule(lr){9-9}\cmidrule(lr){10-10}\cmidrule(lr){11-11}\cmidrule(lr){12-12}

NAC & SSL
& LL-10h
& LTVS-100h & CSJ & COJADS
& WS-100h & AS1
& IEMOCAP & JTES & EmoTalk
& ASR / SER \\
\midrule

\multicolumn{12}{l}{\textbf{Top line (Waveform)}} \\
\midrule
- & Match
& 10.2 / 17.7
& 13.5 & 4.3 & 34.7
& 13.7 / 19.2 & 4.4
& 65.11 $\pm$ 3.1 & 78.47 $\pm$ 4.4 & 77.88
& - \\
\midrule

\multicolumn{12}{l}{\textbf{Effect of SSL Pre-training Language (all models pre-trained on 960h, \Cref{sec:ssl_pretrain_results})}} \\
\midrule
\multirow{3}{*}{All} & EN
& \textbf{12.3 / 22.5}
& 20.4 & 4.8 & 45.1
& 24.4 / 33.3 & 5.5
& \textbf{66.15 $\pm$ 1.9} & 69.30 $\pm$ 5.6 & 66.10
& \textbf{12.49} / 18.61 \\
& JP
& 28.0 / 46.1
& \textbf{15.7} & 4.7 & \textbf{41.6}
& 21.8 / 30.8 & 5.4
& 64.73 $\pm$ 1.6 & \textbf{77.67 $\pm$ 4.2} & 72.10
& 37.89 / \textbf{10.19} \\
& ZH
& 27.5 / 45.7
& 18.4 & \textbf{4.2} & 42.0
& \textbf{16.9 / 24.3} & \textbf{4.8}
& 65.16 $\pm$ 2.2 & 70.56 $\pm$ 6.2 & \textbf{74.79}
& 42.99 / 13.00 \\
\midrule

\multicolumn{12}{l}{\textbf{Effect of NAC Training Language (\Cref{sec:cssl_nac})}} \\
\midrule
EN+ & \multirow{4}{*}{Match}
& \textbf{10.2} / 20.5
& 14.5 & \textbf{4.2} & \textbf{37.5}
& 14.5 / 19.8 & 4.7
& 65.21 $\pm$ 2.5 & 77.92 $\pm$ 3.7 & 77.25
& \textbf{2.47 / 1.38} \\
JP &
& 10.8 / 21.0
& 14.6 & 4.3 & \textbf{37.5}
& \textbf{14.4} / 20.0 & 4.7
& 65.55 $\pm$ 2.7 & 76.72 $\pm$ 4.8 & \textbf{77.41}
& 3.90 / 3.76 \\
ZH &
& 10.4 / \textbf{20.4}
& 14.5 & \textbf{4.2} & 40.2
& \textbf{14.4} / 19.5 & \textbf{4.6}
& \textbf{66.35 $\pm$ 1.9} & 77.12 $\pm$ 3.9 & 75.46
& 2.57 / 5.77 \\
All &
& 10.5 / 21.3
& \textbf{14.2} & 4.8 & 38.8
& \textbf{14.4 / 19.1} & 4.7
& 64.04 $\pm$ 0.9 & \textbf{77.42 $\pm$ 4.6} & 76.14
& 3.06 / 1.87 \\
\bottomrule
\end{tabular}%
}

\end{table*}

\section{Language Sensitivity in\\Codec-Based SSLs} \label{sec:ssl_pretrain}

\subsection{Experimental Setup} \label{sec:ssl}
We now analyze the language sensitivity within codec-based SSLs to answer RQ2 and RQ3.
Unlike RQ1, which used reconstructed waveforms, codec-based SSL uses discrete tokens as inputs.

\noindent\textbf{Datasets.}
SSL pre-training used the language-specific corpora in Table~\ref{tab:data_overview}.
For RQ2, 960 hours of Japanese and Chinese speech were randomly sampled to match English.

\noindent\textbf{SSL pre-training.}
We adopted HuBERT~\cite{hsu2021hubert} implemented in fairseq~\cite{ott2019fairseq}, following the prior codec-based SSL studies~\cite{shi2024espnet,tseng2025codec2vec}.
For RQ2, we followed HuBERT’s two-stage clustering, where stage $0$ was trained for $250$ epochs using $k$-means clusters from MFCC features, and stage 1 for $400$ epochs using clusters from 6th-layer representations, restricting language-dependent factors to the pre-training data.
For RQ3, pseudo labels were generated from language-specific HuBERT Base checkpoints using 9th-layer features and $k$-means clustering.\footnote{\url{https://github.com/facebookresearch/fairseq/tree/main/examples/hubert} for English; \url{https://huggingface.co/imprt/kushinada-hubert-base} for Japanese; \url{https://github.com/TencentGameMate/chinese_speech_pretrain} for Chinese.}
Training followed the official HuBERT recipe\footnote{\url{https://github.com/facebookresearch/fairseq/blob/main/examples/hubert/config/pretrain/hubert_base_librispeech.yaml}} with a batch size of \SI{700}{\second}, \num{8000} warmup updates, learning rate \num{0.001}, and weight decay \num{0.015}.

\noindent\textbf{Waveform and Codec-based SSL Configurations.}
Two input representations were considered under the same HuBERT training framework~\cite{hsu2021hubert}.
\emph{Waveform-input HuBERT} used the standard convolutional feature extractor.
For English, we used the fairseq HuBERT Base model; for Japanese and Chinese, HuBERT Base was pre-trained from scratch with the datasets following the RQ3 recipe.
\emph{NAC-HuBERT} replaced the HuBERT input with discrete tokens produced by a \emph{frozen} NAC.
In the original design of Codec2Vec~\cite{tseng2025codec2vec}, at each step, code indices from all codebooks were mapped to their corresponding pre-trained NAC codebook embeddings and summed to form token embeddings which were fed into the Transformer encoder. In our experiments, we used a frozen NAC quantizer and reused its pre-trained codebook embeddings, rather than learning token embeddings during SSL pre-training.
The Transformer architecture and training objective were the same as those of the HuBERT that takes waveform as input.

\subsection{Effect of Choosing Language in SSL Pre-training} \label{sec:ssl_pretrain_results}
We address RQ2 by varying the SSL pre-training language while fixing the NAC.
The second block of \Cref{tab:rq2_rq3_results} shows the downstream ASR and SER results that answer RQ2.
We used the DAC trained on the All condition, as it showed stable cross-lingual performance in \Cref{sec:comp_nac_train_lang}.
Across languages, aligning the SSL pre-training language to the downstream language generally yielded the best performance for both ASR and SER, with only minor exceptions.
We also observed this tendency under the MFCC-based two-stage clustering (Section~\ref{sec:ssl}), 
indicating that the benefit of aligning languages of the downstream task and the SSL pre-training data is larger than the benefit of using pseudo-labels. 

These results indicate that language-specific information remains after discretization using NAC, and that downstream-relevant structure is mainly acquired during SSL pre-training rather than at the NAC training. 
The better performance when using the same languages across datasets with different recording conditions suggests that the effect is driven primarily by the language differences rather than the domain differences.

\smallskip
\noindent \textbf{RQ2:}
Language sensitivity remains in codec-based SSL even after NAC discretization, implying that cross-lingual differences are governed mainly by the SSL pre-training stage rather than by the NAC training.

\subsection{Effect of Choosing Language in NAC Training with Fixed SSL}
\label{sec:cssl_nac}
To address RQ3, we isolate the effect of NAC training language in codec-based SSL by fixing the SSL pre-training language to match the downstream language and varying only the NAC.
We used DACs trained under four language conditions (EN+, JP, ZH, and All) as described in \Cref{sec:comp_nac_train_lang}.
For each downstream language, we pre-trained NAC-HuBERT on matched-language data and evaluated the ASR/SER performances.

The third block of \Cref{tab:rq2_rq3_results} presents the downstream performances across DAC training-language conditions while keeping the SSL pre-training language fixed.
Performances among DAC training languages were similar, and they did not show a consistent trend, including the All condition.
Notably, using a DAC trained with the same language as the downstream task did not necessarily improve the performance, which is consistent with the CoV results.

Comparing the results of RQ2 and RQ3, particularly for English, suggests that pseudo labels generated by waveform-based HuBERT outperform MFCC-based clustering.
This indicates that pseudo labels by HuBERT with waveform inputs can enhance performances of ASR/SER with codec-based SSLs.

\smallskip
\noindent \textbf{RQ3:}
Once the SSL pre-training language is fixed to the downstream language, retraining an acoustic NAC for each target language is generally unnecessary in codec-based SSL.
This supports reusing a single acoustic NAC across languages without sacrificing downstream performance.
Together with RQ2, the results suggest that language-specific characteristics are primarily acquired during SSL pre-training, while the NAC mainly provides a representation that is less sensitive to language.

\section{Conclusion}
We dissected language sensitivity in codec-based SSL by separating the NAC and SSL pre-training stages.
Across ASR and SER in English, Japanese, and Chinese, acoustic NACs exhibited limited language sensitivity and NAC training language had limited impact, whereas downstream performance was sensitive to SSL pre-training language.
These results support reusing an acoustic NAC across languages and matching SSL pre-training language to the target language.
Future work will extend the analysis to more languages and downstream tasks.

\newpage
\section{Acknowledgments}
This paper is based on results obtained from a project, Programs for Bridging the gap between R\&D and the IDeal society (society 5.0) and Generating Economic and social value (BRIDGE)/Practical Global Research in the AI × Robotics Services, implemented by the Cabinet Office, Government of Japan. This study is also supported by AIST policy-based budget project 'R\&D on Generative AI Foundation Models for the Physical Domain'
\section{Generative AI Use Disclosure}
The authors used ChatGPT to improve the language and clarity of this manuscript.
The authors reviewed and edited the output as needed and take full responsibility for the content of the manuscript.

\bibliographystyle{IEEEtran}
\bibliography{mybib}

\end{document}